\documentclass[12pt]{iopart}

\usepackage{lineno} 
\usepackage{graphicx}
\usepackage{upgreek}

\begin{document}

\title[Constraints on UHECR accelerators]{Observational constraints on accelerators of ultra-high energy cosmic rays}

\author{Sullivan Marafico\textsuperscript{a}, Jonathan Biteau\textsuperscript{a}, Antonio Condorelli\textsuperscript{a}, Olivier Deligny\textsuperscript{a,*}, Quentin Luce\textsuperscript{b}}

\address{\bf \textsuperscript{a} Université Paris-Saclay, CNRS/IN2P3, IJCLab, 91405 Orsay, France\\
\bf \textsuperscript{b} Karlsruhe Institute of Technology, Institute for Experimental Particle Physics (ETP),
Karlsruhe, Germany}
\ead{\bf \textsuperscript{*} deligny@ijclab.in2p3.fr}
\vspace{10pt}
\begin{indented}
\item[]September 2022
\end{indented}

\begin{abstract}
We explore two generic hypotheses for tracing the sources of ultra-high energy cosmic rays (UHECRs) in the Universe: star formation rate density or stellar mass density. For each scenario, we infer a set of constraints for the emission mechanisms in the accelerators, for their energetics and for the abundances of elements at escape from their environments. From these constraints, we generate sky maps above 40~EeV expected from a catalog that comprises 410,761 galaxies out to 350~Mpc and provides a near-infrared flux-limited sample to map both stellar mass and star formation rate over the full sky. Considering a scenario of intermittent sources hosted in every galaxy, we show that the main features observed in arrival directions of UHECRs can in turn constrain the burst rate of the sources provided that magnetic-horizon effects are at play in clusters of galaxies.
\end{abstract}

%
%
%
%
%

\textit{Introduction.} A correlation between arrival directions of ultra-high energy cosmic rays (UHECRs) and the flux patterns of star-forming galaxies within a few Mpc provides currently the most promising evidence for anisotropy above 40~EeV~\cite{PierreAuger:2018qvk,TelescopeArray:2021gxg,PierreAuger:2022axr}. Its meaning, however, remains to be uncovered. In particular, the relatively low-signal strength, on the order of 10--15\%, prevents one from drawing firm statements about an origin of UHECRs exclusively from star-forming galaxies. The aim of this contribution, based on~\cite{Luce:2022awd,Marafico:2022}, is to constrain a benchmark scenario that could accommodate the observed correlation.

Starburst galaxies are responsible for $\simeq 15\%$ of the total star formation rate (SFR) for redshifts $z < 2$~\cite{Rodighiero:2011px,Sargent:2012rj}. If UHECR sources are related to cataclysmic events associated with the deaths of short-lived, massive stars, the rate of which is traced by the SFR, the signal strength is then suggestive of sources being hosted in every star-forming galaxy in the Universe, with starburst galaxies harboring more events than main-sequence galaxies. We explore such a benchmark scenario by considering that the density of sources is fairly traced by the SFR density (SFRD) or, alternatively, by the stellar mass density (SMD). In both cases, we first infer constraints on the emission processes, energetics, and abundances of elements in the source environments by fitting, to the energy spectrum and mass-composition data reported in~\cite{PierreAuger:2021hun,Bellido:2017cgf}, a benchmark scenario in which the intensity of each individual nuclear components is assumed to drop off at the same magnetic rigidity. This is consistent with the basic expectation that electromagnetic processes accelerate particles up to a maximum energy proportional to their electric charge $Z$. Based on these results, we map the UHECR intensity in the sky above 40~EeV that would result from sources following the SFRD or the SMD by using the most complete flux-limited catalog of galaxies over the entire sky~\cite{2021ApJS..256...15B, Biteau:2021bsp}. Considering an intermittent nature for the UHECR sources then enables us to match the arrival directions of UHECRs for some viable range of rate density. 

\begin{figure}[!t]
  \centering
  \includegraphics[width=0.45\textwidth]{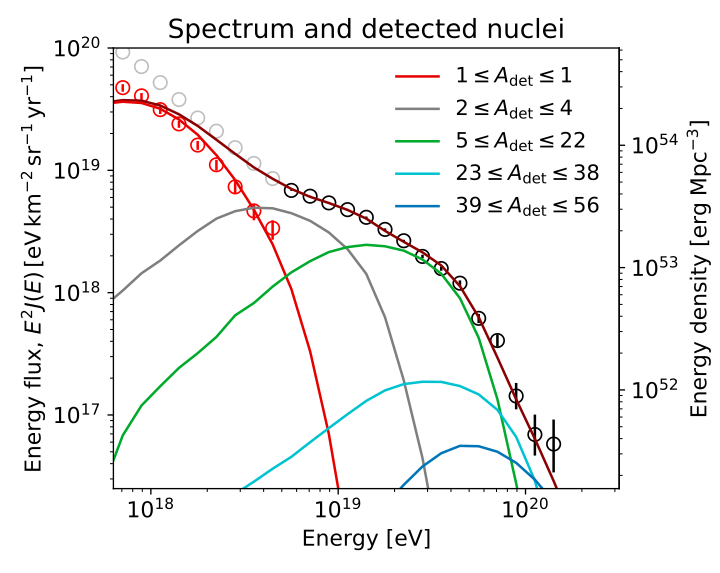}
  \includegraphics[width=0.45\textwidth]{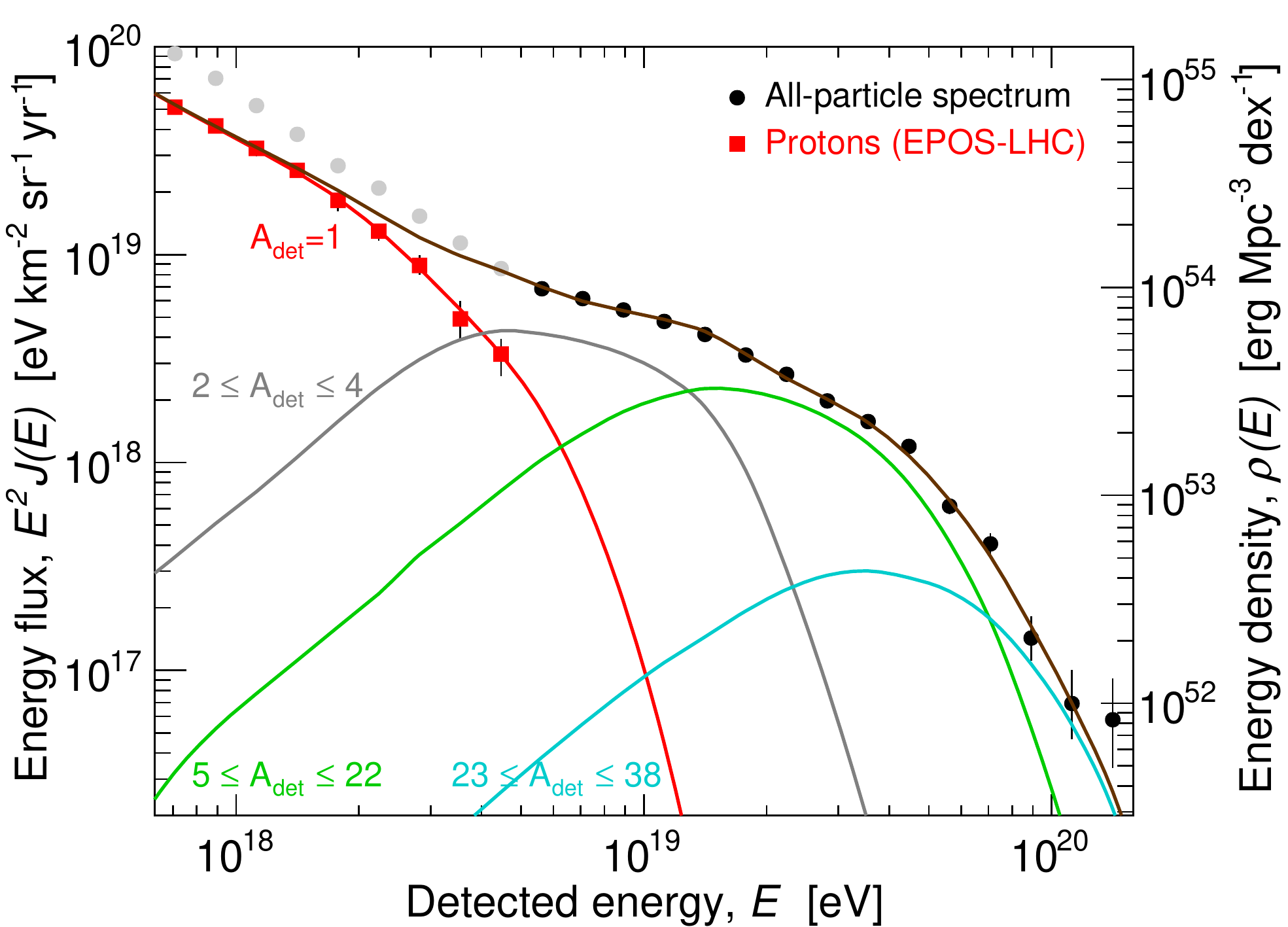}
  \caption{UHECR energy flux on Earth as modeled by the best-fit parameters for the benchmark scenario, for five detected mass groups and for all particles. Left: SMD~\cite{Luce:2022awd}. Right: SFRD~\cite{Marafico:2022}.} 
  \label{fig:e2}
\end{figure}
 
\textit{Source properties required to match observations.} The benchmark model follows here from~\cite{Luce:2022awd}, in which the energy spectrum of the ejected particles as well as the amount of ejected nuclei $A$ may differ strongly from those injected into the electromagnetic field due to in-source interactions, see e.g.~\cite{Unger:2015laa,Globus:2015xga}. This has two important consequences: a) the ejected spectrum of the charged nuclei in $E^{-\gamma_A}$ can be much harder than that injected, due to the escape mechanism; b) the interactions can produce a copious flux of secondary neutrons of energy $E_{\rm n} = E/A$. These neutrons can escape freely from the magnetic confinement zones, with an ejection spectrum  $E^{-\gamma_{\rm p}}$ much softer than that of nuclei, to decay into protons on their way to the Earth. Consequently, the ejection rate of the sources per comoving volume unit and per energy unit of elements is modeled by an exponentially-suppressed power law, the spectral index of which is possibly different for protons ($\gamma_{\rm p}$) and nuclei ($\gamma_A$). 

For reproducing simultaneously the energy spectrum and mass-composition data above $\simeq 0.6$~EeV, the modelled abundance of nuclear elements is found to be dominated by intermediate-mass ones, ranging from He to Si, accelerated to $E^{Z}_{\mathrm{max}}\simeq 5Z\,\mathrm{EeV}$ and escaping from the source environments with a very hard spectral index $\gamma_A$, consistent with previous studies above 5~EeV~\cite{PierreAuger:2016use}. Such a hard spectral index is called for to account for the quasi-monoelemental increase of the mass with energy. In the SMD scenario, fitting the model to the data indeed requires the spectral index for protons ($\gamma_{\rm p}\simeq 3.3$) to be much softer than that of nuclei ($\gamma_{A}\simeq -0.45$)~\cite{Luce:2022awd}. The results are illustrated in the left panel of Fig.~\ref{fig:e2}. In the SFRD scenario, however, the large increase of star-forming activity up to the cosmic noon ($z \simeq 2$) results in a large increase of secondary protons created \textit{en route} subsequently to nuclei interactions. These secondary protons could accommodate the observed spectrum (right panel of Fig.~\ref{fig:e2}), on the condition that $\gamma_{A}\simeq -2$. Inferring a set of constraints on $\gamma_{\rm p}$ and $\gamma_{A}$ would require considering the proton spectrum to even lower energies. 

\textit{UHECR accelerators from Galactic and extragalactic transient events?} If protons of extragalactic origin contribute to the sub-ankle component, other elements are needed to make up the all-particle spectrum shown in Fig.~\ref{fig:e2}.  Interestingly, a steep fall-off of the Fe component above $10^{17}$\,eV has been reported in~\cite{Bellido:2017cgf}. This is along the lines of a scenario for Galactic CRs characterised by a  rigidity-dependent maximum acceleration energy to explain the knee structures. At the same time, the presence of nuclei of the CNO group (and of He nuclei, although with much larger uncertainties) is observed. We advocate here that such intermediate-mass elements could be remnants of an old event producing UHECRs in the Galaxy, in line with the starting assumption that UHECR sources could be harbored in any galaxy. In fact, little is known about the confinement of CRs in the Galaxy by the magnetic field at these energies. Adopting the widely-used value of the confinement volume of the Galactic disk, $\pi (20\,\mathrm{kpc})^2\times 300\,\mathrm{pc}\simeq 10^{67}$\,cm$^3$, intermittent sources such as long gamma-ray bursts releasing $\simeq 10^{51}$\,erg every million years in the Milky Way~\cite{Pohl:2011ve} would supply a luminosity density of a few $10^{-30}$\,erg\,s$^{-1}$\,cm$^{-3}$. For a confinement time of the order of $10^{4}$ or $10^5$\,yr for N nuclei between $10^{17}$ and $10^{18}$\,eV~\cite{1998AstL...24..139Z,Kaapa:2021}, such a luminosity density can accommodate the missing energetics between the all-particle and proton spectra in Fig.~\ref{fig:e2} below the ankle energy. Within this scenario, in which any galaxy could host transient accelerators, UHECRs from the Galactic event would have escaped so that those observed above the ankle energy are indeed from extragalactic events -- see e.g.~\cite{Katz:2013ooa,Loeb:2002ee} for such a scenario. We explore below the constraints that can be obtained on the luminosity required for the UHECR sources by finding the best match between the observed and the expected sky map at the highest energies.

\textit{Steady images of intermittent sources.} Transient events from distant sources can appear as persistent ones at ultra-high energies because of the temporal spread induced by intervening magnetic fields (see, e.g.~\cite{Sigl:2000vf} for a review). The source image of a burst of duration $\delta t$ is then visible during a period of time $\Delta \tau \gg \delta t$ , which can be shown to scale with the amplitude of the magnetic field $B$, the particle rigidity $R$, the source distance $D$ and the coherence length of the field $\lambda_B$ as~\cite{Achterberg1999IntergalacticRays}
\begin{equation}
\label{Eq:time_spread}
   c\Delta \tau \simeq 4.4 \times 10^3 \left( \frac{B}{\rm 10\, nG} \right)^2 \; \left( \frac{R}{10 \, \rm EV} \right)^{-2} \; \left( \frac{D}{1 \rm \, Mpc} \right)^2 \; \left( \frac{\lambda_B}{10 \rm \, kpc} \right) \; {\rm yr}.
\end{equation}
The extragalactic magnetic-field amplitude is currently bracketed between $O(0.1\,$nG) and $O(0.1\,$fG) \cite{2021Univ....7..223A, Bray2018AnAnisotropy, Vazza2017SimulationsObservables}. Here, we consider an intermediate value $B_{\rm XG} = 0.1\,$pG. The time spread induced by such a field is sizeable for sources at cosmological distances, but remains negligible for sources within the GZK horizon. The values for the turbulent component of the Galactic magnetic field are taken from one of the most up-to-date model~\cite{Jansson2012AField}. While this field dominates the angular deflections of UHECRs, its spatial extension is too limited to imprint sizeable time spreads for extragalactic sources. Finally, the last magnetic field considered here is that, largely under-constrained, from the Local Sheet -- referring to the structure of the cosmic web that embeds the Local Group of galaxies. We consider in this exploratory study $B_{\rm LS}$ between 10 and 25~nG, and $\lambda_{\rm B_{\rm LS}} =10{\, \rm kpc}$, extending over 1~Mpc. Such values are consistent with those inferred from MHD simulations~\cite{Donnert2018MagneticSimulation}, although these simulations are highly seed-dependent. 
With these values, the Local Sheet field is the dominant one that governs the time spread of UHECRs, amounting to more than 100,000 years for sources beyond 1~Mpc.

\textit{The sky of transient UHECR sources.} To extract randomly UHECR sources, we use the catalog of galaxies established in~\cite{2021ApJS..256...15B} that comprises 410,761 galaxies out to 350~Mpc, distance at which 50\% of the stellar mass ($M_\star$) is below the 2MASS sensitivity limit. This catalog is designed to account for incompleteness as a function of luminosity distance, limited by the 2MASS sensitivity threshold, and of angular distance to the Galactic plane, where source confusion and obscuration mask background galaxies in the so called Zone of Avoidance. It provides an unprecedented view on $M_\star$ and SFR on scales ranging from the very outskirts of the Milky Way out to the most distant superclusters identified in Cosmic Flows~\cite{Hoffman:2017ako}.

\begin{figure}[!t]
  \centering
  \includegraphics[width=0.46\textwidth]{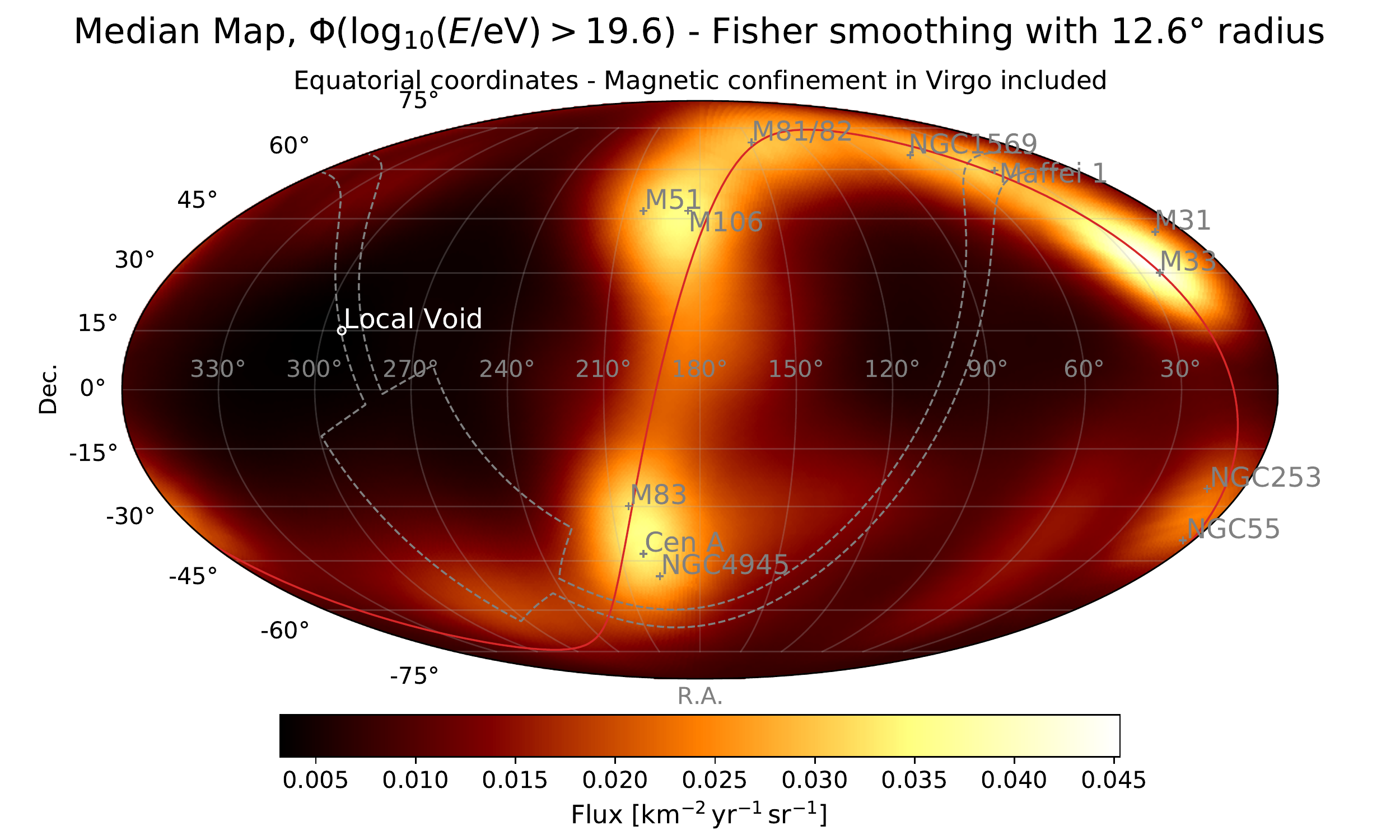}
  \includegraphics[width=0.53\textwidth]{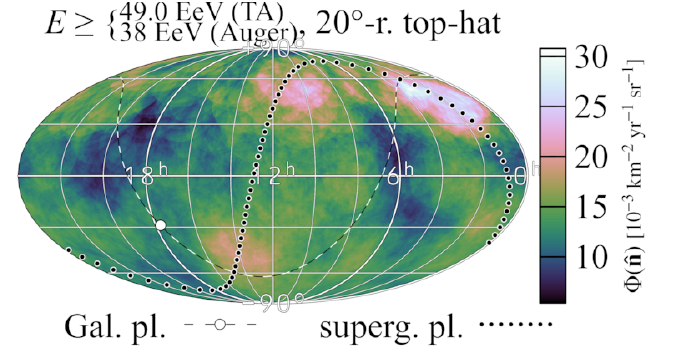}
  \caption{Left: Median sky map in the scenario of intermittent sources traced by the SFRD, using a 12.6-degree gaussian smearing, for $k=2~10^{-5}$~$\textup{M}_\odot^{-1}$~\cite{Marafico:2022}. Right: Observed arrival directions~\cite{TelescopeArray:2021gxg}.} 
  \label{fig:maps}
\end{figure}

The Virgo cluster, at a distance of $\simeq 16~$Mpc, is the major cluster of galaxies that is within the UHECR horizon above $\simeq 40~$EeV. Accelerators located in the galaxies of this cluster may not, however, contribute to the UHECR intensity observed on Earth due to the longer confinement time scales encountered in this environment. Magnetic-field amplitudes on the order of several $\upmu$G are expected at cluster centers and are indeed observed in the Coma cluster (central value of $4.7\pm0.7~\upmu$G)~\cite{Bonafede:2010xg}. A similar amplitude is expected in Virgo based on its pressure profile inferred from Planck and X-ray observations \cite{2016A&A...596A.101P}. Under the approximation of a coherence length and extent similar to those of Coma, the $B$-field in the Virgo-cluster environment is then sufficient to confine cosmic rays longer than their energy-loss time for the particle rigidities of interest. Consequently, we consider in this exploratory study that UHECRs from galaxies in the Virgo cluster could never reach Earth. A more detailed and careful treatment of the magnetic confinement in galaxy clusters is underway~\cite{Condorelli:2022}. Note also that propagation effects could prevents UHECRs from far-enough distant sources located in the Local Sheet to reach us~\cite{Sigl:1998dd}.

In the scenario of intermittent sources, the probability to observe $N$ bursting sources from a given host galaxy follows a Poisson distribution, with $N=ks\Delta\tau$. The rate of bursting sources is governed by the parameter $k$, expressed in $\textup{M}_\odot^{-1}$ units for $s$ tracing the SFRD or in $\textup{M}_\odot^{-1}~\mathrm{yr}^{-1}$ units for $s$ tracing the SMD. Using the best-fit parameters obtained from the combined fit of the energy spectrum and mass-composition data, the flux from each direction in the sky can be evaluated by sampling randomly the Poisson process for a given value of $k$. As a fair illustration of the various different maps that result from the stochastic process, we show in the left panel of Fig.~\ref{fig:maps} the median map for $k=2~10^{-5}$~$\textup{M}_\odot^{-1}$ in the SFRD scenario above 40~EeV, using a 12.6-degree gaussian smearing (that corresponds to a 20-degree top hat smearing). The model displays overdensities around the supergalactic plane, as expected from the galaxies in the Local Sheet that are inferred to be responsible for the indication of anisotropies at the $4.2\sigma$ confidence level at these energies (right panel)~\cite{TelescopeArray:2021gxg}. Similar qualitative agreement can be found in the SMD case for $k\simeq 10^{-15}$~$\textup{M}_\odot^{-1}~\mathrm{yr}^{-1}$.

\begin{figure}[!t]
  \centering
  \includegraphics[width=0.45\textwidth]{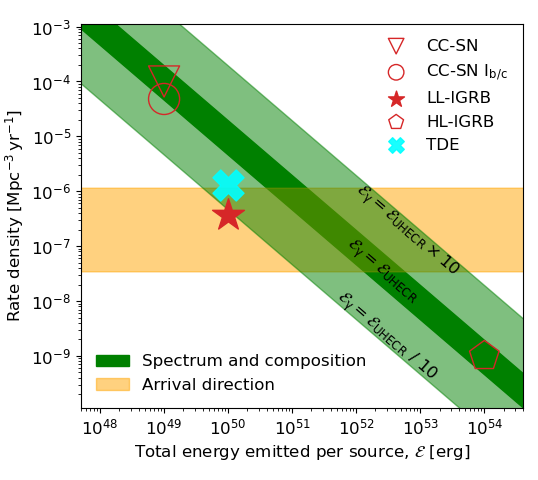}
  \caption{Constraints on the density of UHECR source rate as a function of the total energy emitted per source, with a few putative sources (CC-SN ($I_{\mathrm{b/c}}$): core-collapse supernovae (of type $I_\mathrm{b/c}$), L(H)L-lGRB: low (high) luminosity long gamma ray bursts, TDE: tidal disruption events.) From~\cite{Marafico:2022}.} 
  \label{fig:diagram}
\end{figure}

\textit{Constraints on UHECR accelerators.} We finally compare our results to the rate density and the total energy emitted in photons of a few source candidate in Fig.~\ref{fig:diagram}. The dark green band in diagonal corresponds to the UHECR luminosity obtained in this work for both SMD and SFRD scenarios, indistinctly. In the absence of firmly-prescribed correspondences between UHECR and photon luminosity, the lighter green band extends to a factor 10 the putative proportionality. All five selected source classes are observed within the green region. The yellow band, on the other hand, results from the constraints obtained on $k$. Objects such as low-luminosity long gamma ray bursts (LL-lGRB) in the SFRD case and events such as tidal disrupted ones (TDE) in the SMD case are observed to match the two sets of constraints. Interestingly, their energetics and source rate lie in the range of the fiducial values needed to fill the sub-ankle energy flux with a Galactic event.
 
In deriving the constraints, the role of magnetic fields in galaxy clusters and in the Local Sheet is key. The understanding of these fields will benefit from X-ray and radio observations with eROSITA and SKA and will allow for deciding whether the scenario of UHECR transient sources harbored in any galaxy is viable or not. \\

\bibliographystyle{iopart-num}
\bibliography{biblio}
 
\end{document}